\documentclass[twocolumn,showpacs,showkeys,preprintnumbers,amsmath,amssymb,prd]{revtex4}
\usepackage{mathptmx}
\usepackage{graphicx}
\usepackage{dcolumn}
\usepackage{bm}

\begin{document}
\preprint{HUPD-0404}
\title{Effective Potential for $\lambda\phi^4$ Theory
at Finite Temperature in $R\otimes S^{D-1}$ and $R\otimes H^{D-1}$
}

\author{T.~Hattori}
\author{M.~Hayashi}
\author{T.~Inagaki}%
 \email{inagaki@hiroshima-u.ac.jp}
\altaffiliation[Also at ]{%
Information Media Center, Hiroshima University
}%
\author{Y.~Kitadono}
\affiliation{Department of Physics, Hiroshima University.
Higashi-Hiroshima, Hiroshima, 739-8521, Japan
}

\date{\today}

\begin{abstract}
We calculate the explicit expression of the effective potential in 
a $\lambda\phi^4$ theory at finite temperature in a static universe
for arbitrary spacetime dimensions ($2\leq D < 5$).
To study the combined effects of the temperature and scale factor to 
the spontaneous symmetry breaking we evaluate the effective potential 
at finite temperature in $R\otimes S^{D-1}$ and $R\otimes H^{D-1}$.
The phase structure of a $\lambda\phi^4$ theory is found by observing the
minimum of the effective potential with varying temperature and scale 
factor. All the ring diagrams are resummed for $D\gtrsim 4$ to 
improve the loop expansion at high temperature. For a conformally 
coupled and a minimally coupled scalar field it is shown 
that temperature and positive curvature suppress the symmetry breaking, 
while negative curvature enhances it. The conformally coupled scalar has
larger curvature effects than the minimally coupled one.
\end{abstract}
\maketitle

\section{Introduction}
The idea of spontaneous symmetry breaking has an important role in 
elementary particle physics. In fact it is understood that the 
electroweak symmetry is spontaneously broken down through the Higgs
mechanism. Grand unified theories (GUT) are constructed on a basis of 
spontaneous gauge symmetry breaking to yield a theory at low energy 
scale. It is generally expected that a more fundamental theory with a 
higher symmetry is realized in the early universe. Thus one of the 
possible environment to test the mechanism of symmetry breaking is 
found in the early universe where the fundamental symmetry is broken 
down. In the early stage of universe, especially at GUT era, we can
neglect the thermal and curvature effects. The spontaneous symmetry 
breaking takes place under the influence of the high temperature, 
strong curvature and non-trivial topology.  

Much attention has been paid in the study of the spontaneous 
symmetry breaking under the circumstance of the high temperature, 
strong curvature and non-trivial topology. For this purpose
finite temperature quantum field theory is investigated in curved
space time \cite{DC, AD, BA, AS, Ken, Gusev:1998rp, Frenkel:1994bg, %
Kirsten:1992vq, Hu:1982ue, Chen:1985wb, Hu:1986jh, Cognola:1992jt, %
Fursaev:1993hm, Roy:1989gj}
The effective action of the system is calculated to determine the 
ground state. Where the ground state of the system breaks a symmetry 
of the Lagrangian, the symmetry is broken down spontaneously. 
If the universe is almost static, homogeneous and isotropic, it is 
quite natural to assume an order parameter of the symmetry dose not
depend on the spacetime coordinates. Under this assumption a ground 
state is found by observing the minimum of the effective potential 
as a function of an order parameter \cite{Hu:1982ue}. 
Hence, the effective potential of many kinds of quantum field theories 
is evaluated at finite temperature and/or finite curvature with the 
spacetime structure fixed (for a review, see Ref. \cite{BOS} and 
references therein).

In the present paper we consider a real scalar fields with $\phi^4$ 
interaction in static homogeneous and isotropic spacetime. 
The $\lambda\phi^4$ theory is one of the simplest models where a 
discrete symmetry is spontaneously broken down. To find the ground 
state of the system we evaluate the effective potential at high 
temperature and strong curavture. 
In two, three and four spacetime dimensions the theory is renormalizable. 
Beyond five dimensions the theory is nonrenormalizable. We confine 
ourselves to the spacetime dimensions greater than or equal to 2 and 
less than 5 and calculate the renormalized effective potential.
The curvature effect to the symmetry breaking comes from a scale
factor dependence of the covariant derivative and a coupling between 
the scalar field and the gravitational field.
One-loop and two-loop corrections to scalar field theories in
linear curvature approximation were found in Refs. \cite{BO1}
and \cite{BO2}, respectvely. This method is gneral, any renormalizable 
theory maybe studied in this way (including backgrounds of non-trivial 
topology) but at small curvaure only.

We assume that the system is in equilibrium and introduce the 
temperature. The assumption is not acceptable in a general curved 
spacetime. We then restrict ourselves in the static universe. Therefore 
we work in the scalar $\lambda\phi^4$ theory in the positive curvature 
spacetime $R\otimes S^{D-1}$ and negative curvature spacetime 
$R\otimes H^{D-1}$. An exact expression for scalar and spinor two-point 
functions is calculated in $R\otimes S^{D-1}$ and $R\otimes H^{D-1}$
\cite{CR, AJ, AL, C, IIM0, IIM1}. It is one of the fundamental object
in dealing with quantum field theories. The vacuum energy density for 
scalar and spinor fields is obtained in $R\otimes S^{D-1}$
\cite{JD,LF,DA}. Thermal effect to the vacuum energy is discussed 
in \cite{DC,AD,BA,AS}.
The curvature effect for symmetry breaking is studied for the $\lambda\phi^4$ 
theory and the four-fermion theory in $R\otimes S^{D-1}$
\cite{O'Connor:1983fq, IIM2, II}.
It has been found that the broken symmetry is restored for a 
sufficiently small scale factor. 
Following the imaginary time formalism, we calculate the effective
potential at finite temperature in curved spacetime.
For $D\gtrsim 4$ a naive coupling expansion is not valid at high temperature. 
To improve the loop expansion we resum all the ring diagrams \cite{FT} 
and estimate the effective potential for $D\gtrsim 4$.

The paper is organized in the following way.
In Sec.II we briefly review a basic formalism to calculate the effective 
potential for a $\lambda\phi^4$ theory in curved spacetime. 
In Sec.III we calculate the effective potential in $R\otimes S^{D-1}$ 
and $R\otimes H^{D-1}$ without making any approximation in the spacetime
curvature. According to the imaginary time formalism the 
temperature is introduced in the theory. The ordinary renormalization
procedure in a flat spacetime is used to obtain the renormalized
effective potential.
In Sec.IV we numerically calculate the effective potential at finite
temperature in $R\otimes S^{D-1}$ and $R\otimes H^{D-1}$.
Characteristic behaviors of the effective potential are shown as
the scale factor varies in the fixed temperature.
Section V gives the concluding remarks.

\section{$\lambda\phi^{4}$ theory in curved spacetime}
In this section we summarise how to obtain the effective potential in the
Feynmann path-integral formalism. Here we consider the $\lambda\phi^{4}$ 
theory constructed by a real scalar field with a $\phi^{4}$ interaction. 
It is one of the simplest models where the spontaneous symmetry breaking 
takes place. The theory is defined by the Lagrangian,
\begin{equation}
  {\cal L}(\phi) = - \frac{1}{2}\phi(\Box + \xi_0 R )\phi
  +\frac{\mu_0^2}{2}\phi^2 
  -\frac{\lambda_0}{4!}\phi^4,
\end{equation}
where $\phi$ is a real scalar field, $i \mu_0$ corresponds to
the bare mass of the scalar field, $\lambda_0$ the bare coupling
constant for the scalar self-interaction, $\xi_0$ the bare coupling 
constant between the scalar field and the gravitational field.

The Lagrangian is invariant under the discrete transformation,
\begin{equation}
  \phi \rightarrow -\phi .
\label{def:disc}
\end{equation}
This $Z_2$ symmetry prevents the Lagrangian from having $\phi^{3}$ terms. 
Because of the minus sign of the mass term the symmetric state,
$\langle \phi\rangle =0$, is unstable in a classical level.
If the non-vanishing expectation value is assigned to the field $\phi$ 
in the ground state, the discrete symmetry is eventually broken.
To see the phase structure of the theory we want to find a ground state.

In quantum theories the ground state is determined by observing the 
minimum of the effective potential. Here we briefly review the effective 
potential for $\lambda\phi^4$ theory. 
We start with the generating functional of the theory which is given by
\begin{equation}
  e^{\frac{i}{\hbar}W[J]}
  \equiv \int {\cal D} \phi \  
  e^{\frac{i}{\hbar}\int d^{D}x \sqrt{-g} \left({\cal L}(\phi)
  +\phi J\right)} , 
\label{GeF}
\end{equation}
where $g$ is the determinant of the metric tensor $g_{\mu\nu}$. 
In the presence of the source $J$ the classical equation of motion becomes 
\begin{equation}
  \left.\frac{\delta S[\phi]}{\delta \phi(x)}\right|_{\phi  = \phi_b} = -J(x),
\label{EoM}
\end{equation}
where $S[\phi]$ is the action
\begin{equation}
  S[\phi] \equiv \int d^{D}x \sqrt{-g} {\cal L}(\phi).
\label{def:S}
\end{equation}
In curved spacetime the functional derivative is defined by
\begin{equation}
  S[\phi+\delta\phi] - S[\phi]
  = \int d^{D}x \sqrt{-g} \frac{\delta S}{\delta \phi(x)} \delta \phi(x).
\end{equation}

We divide the field $\phi$ into a classical background 
$\phi_{b}$ which satisfies the Eq.(\ref{EoM})
and a quantum fluctuation $\tilde{\phi}$,
\begin{equation}
\phi = \phi_{b} + \hbar^{1/2}\tilde{\phi}.
\end{equation}
In terms of $\phi_{b}$ and $\tilde{\phi}$ 
the action $S[\phi]$ is rewritten as
\begin{eqnarray}
  &&S[\phi] = S[\phi_{b} + \hbar^{1/2}\tilde{\phi}]
\nonumber \\
  &&= S[\phi_{b}]
    -\int d^{D}x \sqrt{-g}J(x) \hbar^{1/2}\tilde{\phi}(x)
\nonumber \\
  &&+\frac{\hbar}{2}\int d^{D}x \int d^{D}y \sqrt{-g(x)} \sqrt{-g(y)}
    \tilde{\phi}(x)iG^{-1}(x,y)\tilde{\phi}(y)
\nonumber \\
  && +\mbox{O}({\hbar^{3/2}\tilde{\phi}^3}),
\end{eqnarray}
where $G^{-1}(x,y)$ is the scalar two-point function,
\begin{eqnarray}
  iG^{-1}(x,y) & \equiv &
  \frac{\delta^2 S[\phi]}{\delta \phi(x)\delta\phi(y)}
\nonumber \\
  & = & -  \left(\Box + \xi_0 R + \mu_0^2  
        -\frac{\lambda_0}{2}\phi_b^2 \right) .
\label{def:g}
\end{eqnarray}
Therefore the generating functional (\ref{GeF}) is expanded to be
\begin{eqnarray}
  &&e^{\frac{i}{\hbar}W[J]} = e^{\frac{i}{\hbar}\left(S[\phi_{b}]
    -\int d^{D}x \sqrt{-g}J(x) \phi_b(x) \right)}
\nonumber \\ 
  && \times \int {\cal D} \tilde{\phi}\ e^{\frac{i}{2}
    \int d^{D}x \int d^{D}y \sqrt{-g(x)} \sqrt{-g(y)}
    \tilde{\phi}(x)iG^{-1}(x,y)\tilde{\phi}(y)+\mbox{O}(\hbar^{1/2})}.
\nonumber \\
\end{eqnarray}
Performing the path-integral,
\begin{eqnarray}
  &&\int {\cal D} \tilde{\phi}\ e^{\frac{i}{2}
  \int d^{D}x \int d^{D}y \sqrt{-g(x)} \sqrt{-g(y)}
  \tilde{\phi}(x)iG^{-1}(x,y)\tilde{\phi}(y)}
\nonumber\\
  &&= \left[\mbox{Det}\ iG^{-1}\right]^{-1/2} ,
\end{eqnarray}
we obtain the generating functional
\begin{eqnarray}
  W[J] &=& S[\phi_{b}]
  -\int d^{D}x \sqrt{-g}J(x) \phi_b(x)
\nonumber\\
&&  -\frac{i\hbar}{2}\ln \left[\mbox{Det}\ iG^{-1}\right] 
  +\mbox{O}(\hbar^{3/2}).
\label{def:w}
\end{eqnarray}

The effective action $\Gamma [\phi_c]$ is given by the Legendre transform
of $W[J]$,
\begin{equation}
  \Gamma [\phi_c] \equiv W[J] - \int d^{D}x \sqrt{-g} \phi_c(x)J(x),
\end{equation}
where $\phi_c$ denotes the expectation value of $\phi$ in the
presence of the source $J$,
\begin{equation}
  \phi_c \equiv \frac{\delta W[J]}{\delta J}.
\end{equation}
From Eq. (\ref{def:w}) the effective action is expanded to be
\begin{equation}
  \Gamma [\phi_c] = S[\phi_c]
  -\frac{i\hbar}{2}\ln \left[\mbox{Det}\ iG^{-1}\right] 
  +\mbox{O}(\hbar^{3/2}).
\end{equation}

If there is a translational invariance, $\phi_c(x)$ in the ground state
dose not depend on the spacetime coordinates. In such a case it is more 
convenient to consider the effective potential.
The effective potential $V(\phi)$ is defined by
\begin{eqnarray}
  V (\phi) & = & - \frac{1}{\Omega}\Gamma[\phi_c(x) = \phi]
\nonumber \\
  &=& \frac{\xi_0}{2}R\phi^2 -\frac{\mu_0^2}{2}\phi^2 
    +\frac{\lambda_0}{4!}\phi^4
    +\frac{i\hbar}{2 \Omega}\ln \left[\mbox{Det}\ iG^{-1}\right] 
\nonumber \\
  &&  +\mbox{O}(\hbar^{3/2}) ,
\label{EP}
\end{eqnarray}
where we put a constant value $\phi_c(x)=\phi$ and $\Omega$ is the 
spacetime volume
\begin{equation}
  \Omega = \int d^{D}x \sqrt{-g} .
\end{equation}

The expectation value of $\phi$ in the ground state satisfies
\begin{equation}
  \langle \phi \rangle \equiv \lim_{J\rightarrow 0} \frac{\delta W[J]}{\delta J}.
\end{equation}
This equation is rewritten as 
\begin{equation}
  \left. \frac{\delta \Gamma[\phi_c]}{\delta \phi_c} \right|_{\phi_c \rightarrow \langle \phi \rangle} = 0 ,
\label{Gap:Action}
\end{equation}
where we have used the relation
\begin{equation}
  \frac{\delta \Gamma[\phi_c]}{\delta \phi_c} = J.
\end{equation}
The equation (\ref{Gap:Action}) is called the gap equation. 
Here we assume that the expectation value of $\phi$ in the ground state 
is independent of the spacetime coordinate $x$. In this case the gap equation 
(\ref{Gap:Action}) reduces to
\begin{equation}
  \left. \frac{d V(\phi)}{d \phi} \right|_{\phi \rightarrow \langle \phi \rangle} = 0 .
\label{Gap:V}
\end{equation}
The expectation value $\langle \phi \rangle$ is obtained
by solving the gap equation (\ref{Gap:V}).

\section{Effective potential in $R\otimes S^{D-1}$ and $R\otimes H^{D-1}$}
Here we introduce the curvature and temperature in the theory and calculate 
the effective potential. First we consider the constant curvature space
$R\otimes S^{D-1}$ and $R\otimes H^{D-1}$ as Euclidean analog of the static 
Einstein universe.
The manifold $R\otimes S^{D-1}$ is represented by the metric
\begin{equation}
  ds^2 = dr^2 + a^2 (d\theta^2+\sin^2\theta d\Omega_{D-2}), 
\end{equation}
where $d\Omega_{D-2}$ is the metric on a unit sphere $S^{D-2}$ and
$a$ is the scale factor. It is a constant curvature spacetime with 
positive curvature
\begin{equation}
  R = (D-1)(D-2)\frac{1}{a^2}.
\end{equation}
The manifold $R\otimes H^{D-1}$ which is represented by the metric
\begin{equation}
  ds^2 = dr^2 + a^2 (d\theta^2+\sinh^2\theta d\Omega_{D-2}), 
\end{equation}
is also a constant curvature spacetime with negative curvature
\begin{equation}
  R = -(D-1)(D-2)\frac{1}{a^2}.
\end{equation}

\subsection{Effective potential at $T=0$}
The effective potential (\ref{EP}) is described by the two-point
function $G(x,y)$ of the real scalar fields. The effect of the
spacetime structure is introduced to the effective potential
through this two-point function. 

In Euclidean spacetime the effective potential $V(\phi)$ reads
\begin{eqnarray}
  V (\phi) &=& \frac{\xi_0}{2}R\phi^2 -\frac{\mu_0^2}{2}\phi^2 
    +\frac{\lambda_0}{4!}\phi^4
    +\frac{\hbar}{2 \Omega}\ln \left[\mbox{Det}\ (-G^{-1})\right] 
\nonumber \\
  &&    +\mbox{O}(\hbar^{3/2})
\nonumber \\
  &=& \frac{\xi_0}{2}R\phi^2 -\frac{\mu_0^2}{2}\phi^2 
    +\frac{\lambda_0}{4!}\phi^4
    +\frac{\hbar}{2 \Omega}\mbox{Tr} \ln\ (-G^{-1}) 
\nonumber \\
  &&    +\mbox{O}(\hbar^{3/2}) .
\label{EP0:E}
\end{eqnarray}
The scalar two-point function satisfies the Klein-Gordan equation
\begin{eqnarray}
  && \left[(\partial_4)^2+\Box_{D-1}-\xi_0 R +\mu_0^{2}
  -\frac{\lambda_0}{2}\phi^2\right]G(x,y)
\nonumber \\
  &&  = -\frac{1}{\sqrt{g}}\delta^{D}(x,y) ,
\label{KG}
\end{eqnarray}
where $\Box_{D-1}$ is the Laplacian on $S^{D-1}$.
Thus the effective potential $V(\phi)$ is rewritten as
\begin{eqnarray}
  V (\phi) &=& \frac{\xi_0}{2}R\phi^2 -\frac{\mu_0^2}{2}\phi^2 
    +\frac{\lambda_0}{4!}\phi^4
\nonumber \\
  &&  +\frac{\hbar}{2 \Omega}\mbox{Tr} \ln\ 
    \left[(\partial_4)^2+\Box_{D-1}-\xi_0 R +\mu_0^{2}
    -\frac{\lambda_0}{2}\phi^2\right]
\nonumber \\
  &&    +\mbox{O}(\hbar^{3/2})
\nonumber \\
  &=& \frac{\xi_0}{2}R\phi^2 -\frac{\mu_0^2}{2}\phi^2 
    +\frac{\lambda_0}{4!}\phi^4
\nonumber \\
  &&  -\frac{\hbar \lambda_0}{4 \Omega}
    \int_0^{\phi^2}dm^2\ 
\nonumber \\
  &&\times    \mbox{Tr} 
    \left[(\partial_4)^2+\Box_{D-1}-\xi_0 R + \mu_0^2 
    - \frac{\lambda_0}{2}m^2\right]^{-1}.
\label{EP1:E}
\end{eqnarray}
At the last line we normalise the effective potential so that
$V(0)=0$ and neglect the O$(\hbar^{3/2})$ term. The integrand 
in the last line of the Eq.(\ref{EP1:E}) 
is described by the Green function $G(x,x,m)$,
\begin{eqnarray}
&&    \mbox{Tr} 
    \left[(\partial_4)^2+\Box_{D-1}-\xi_0 R + \mu_0^2
    -\frac{\lambda_0}{2}m^2\right]^{-1}
\nonumber \\
  &&    = -\int d^{D}x \sqrt{g} G(x,x;\phi=m).
\label{TrG}
\end{eqnarray}
Thus the effective potential reads
\begin{eqnarray}
  V (\phi) &=& \frac{\xi_0}{2}R\phi^2 -\frac{\mu_0^2}{2}\phi^2 
    +\frac{\lambda_0}{4!}\phi^4
\nonumber \\
  &&    +\frac{\hbar\lambda_0}{4}
    \int_0^{\phi^2}dm^2\ G(x,x;m).
\label{EP:E}
\end{eqnarray}
Substituting Eq.(\ref{EP:E}) to Eq.(\ref{Gap:V}),
we obtain the following expression for the gap equation
\begin{eqnarray}
  &&\left. \frac{\delta V(\phi)}{\delta \phi} \right|_{\phi \rightarrow \langle \phi \rangle} 
\nonumber \\
  &&  = \langle\phi\rangle
    \left(\xi_0 R - \mu_0^2 + \frac{\lambda_0}{2}\langle \phi \rangle^2
    +\frac{\hbar\lambda_0}{2} G(x,x;\langle \phi \rangle)\right)
  = 0.
\nonumber \\
\label{Gap:E}
\end{eqnarray}

On the manifolds $R\otimes S^{D-1}$
the solution of Eq.(\ref{KG}) is given by\cite{CR, AJ, AL, C, IIM0, IIM1}
\begin{eqnarray}
  &&G(x,y;m)=\frac{a^{3-D}}{(4\pi)^{(D-1)/2}}
    \int \frac{d\omega}{2\pi} e^{-i\omega (y-x)_{4}}
\nonumber \\
  && \times
      \frac{\displaystyle \Gamma\left(\frac{D-2}{2}+i\alpha_S\right)
                        \Gamma\left(\frac{D-2}{2}-i\alpha_S\right)}
         {\displaystyle \Gamma\left(\frac{D-1}{2}\right)}
\label{sctf:S} \\
  && \times
    F\left(\frac{D-2}{2}+i\alpha_S, \frac{D-2}{2}-i\alpha_S,
            \frac{D-1}{2};\cos^{2}\left(\frac{\sigma}{2a}\right)\right) ,
\nonumber
\end{eqnarray}
where $\sigma$ is the geodesic distance between $x$ and $y$ on $S^{D-1}$
and $\alpha_S$ is defined by
\begin{equation}
  \alpha_S \equiv \sqrt{f(\omega)a^{2}
             +(D-1)(D-2)\xi_0
             -\frac{(D-2)^{2}}{4}}\, ,
\end{equation}
with
\begin{equation}
f(\omega)\equiv\omega^{2}-\mu_0^{2}+\frac{\lambda_0}{2}m^2 .
\label{def:fomega}
\end{equation}
The two-point function (\ref{sctf:S}) develops many singularities at
$\sigma=2\pi n a$ where $n$ is an arbitrary integer.
This property is a direct consequence of the boundedness of the
manifold $S^{D-1}$.
In other words the geodesic distance $\sigma$ is bounded in $[0,2\pi a)$.
Thus the two-point function (\ref{sctf:S}) satisfies the periodic boundary
condition $G((y-x)_{4},\sigma)=G((y-x)_{4},\sigma+2\pi n a)$.

Following the procedure developed in Ref. \cite{IIM0} we can solve 
Eq.(\ref{KG}) on the manifold $R\otimes H^{D-1}$ and find the 
two-point function,
\begin{eqnarray}
  &&G(x,y;m)=\frac{a^{3-D}}{(4\pi)^{(D-1)/2}}
     \int \frac{d\omega}{2\pi} e^{-i\omega (y-x)_{4}}
\nonumber \\
  && \times
         \cosh^{2-D-\alpha_H}\left(\frac{\sigma}{2a}\right)
          \frac{\displaystyle \Gamma\left(\frac{D-2}{2}+\alpha_H\right)
                          \Gamma\left(\frac{1}{2}+\alpha_H\right)}
           {\displaystyle \Gamma\left(2\alpha_H +1 \right)}
\nonumber\\
  && \times
    F\left(\frac{D-2}{2}+\alpha_H, \frac{1}{2}+\alpha_H,
           2\alpha_H +1;\cosh^{-2}\left(\frac{\sigma}{2a}\right)\right) ,
\nonumber\\
\label{sctf:H} 
\end{eqnarray}
where $\sigma$ is the geodesic distance between $x$ and $y$ on $H^{D-1}$
and $\alpha_H$ is defined by
\begin{equation}
  \alpha_H \equiv \sqrt{f(\omega)a^{2}
            -(D-1)(D-2)\xi_0
            +\frac{(D-2)^{2}}{4}}\, .
\end{equation}
On $R\otimes H^{D-1}$ the geodesic distance $\sigma$ is not bounded.
The two-point function (\ref{sctf:H}) dose not have any singularities 
except for the limit $\sigma\rightarrow 0$.

Substituting Eqs.(\ref{sctf:S}) and (\ref{sctf:H}) to Eq.(\ref{EP:E}), 
we obtain the effective potential $V(\phi)$ on $R\otimes S^{D-1}$,
\begin{eqnarray}
  && V (\phi)  =  \frac{\xi_0}{2}R\phi^2 -\frac{\mu_0^2}{2}\phi^2 
    +\frac{\lambda_0}{4!}\phi^4
\nonumber\\
   && +\frac{\hbar\lambda_0}{4}\frac{a^{3-D}}{(4\pi)^{(D-1)/2}}
     \Gamma\left(\frac{3-D}{2}\right)
\label{EP:RS} \\
  && \times \int_0^{\phi^2}dm^2\int\frac{d\omega}{2\pi}
     \frac{\displaystyle \Gamma\left(\frac{D-2}{2}+i\alpha_S\right)
                    \Gamma\left(\frac{D-2}{2}-i\alpha_S\right)}
          {\displaystyle \Gamma\left(\frac{1}{2}+i\alpha_S\right)
                    \Gamma\left(\frac{1}{2}-i\alpha_S\right)},
\nonumber
\end{eqnarray}
and on $R\otimes H^{D-1}$,
\begin{eqnarray}
   V (\phi) &=& \frac{\xi_0}{2}R\phi^2 -\frac{\mu_0^2}{2}\phi^2 
    +\frac{\lambda_0}{4!}\phi^4
\nonumber\\
   &&    +\frac{\hbar\lambda_0}{4}\frac{a^{3-D}}{(4\pi)^{(D-1)/2}}
     \Gamma\left(\frac{3-D}{2}\right)
\nonumber \\
  && \times \int_0^{\phi^2}dm^2\int\frac{d\omega}{2\pi}
     \frac{\displaystyle \Gamma\left(\frac{D-2}{2}+\alpha_H\right)}
          {\displaystyle \Gamma\left(\frac{4-D}{2}+\alpha_H\right)},
\label{EP:RH}
\end{eqnarray}
where we have used the relation
\begin{equation}
  F(\alpha, \beta, \gamma, 1)=
  \frac{\displaystyle \Gamma(\gamma) \Gamma(\gamma-\alpha-\beta)}
       {\displaystyle \Gamma(\gamma-\alpha) \Gamma(\gamma-\beta)}.
\end{equation}

\subsection{Effective potential at finite temperature}
Next we introduce the effect of the finite temperature.
Since the manifolds $R\otimes S^{D-1}$ and $R\otimes H^{D-1}$
have no time evolution, the equilibrium state can be defined.
Here we follow the standard procedure of the imaginary time 
formalism.

At finite temperature Eq.(\ref{TrG}) is modified as
\begin{eqnarray}
  &&  \mbox{Tr} 
      \left[(\partial_4)^2+\Box_{D-1}-\xi_0 R + \mu_0^2
      -\frac{\lambda_0}{2}m^2\right]^{-1}
\nonumber \\
  &&  = -\int_0^{\beta} dx_{4} \int d^{D-1}x \sqrt{g} G(x,x;\phi=m),
\label{TrGT}
\end{eqnarray}
where $\beta=1/(k_BT)$ with $k_B$ the Boltzmann constant and $T$
the temperature. The fourth component of the coordinate $x_4$ is
bounded in $[0,\beta)$.

Following the standard procedure of the imaginary time formalism 
(for a review, see Ref. \cite{FT}),
the scalar two point function at finite temperature is obtained 
by the replacement, 
\begin{eqnarray}
  \left\{
  \begin{array}{l}
    \displaystyle
    \int\frac{d\omega}{2\pi} \rightarrow 
    \frac{1}{\beta}\sum_{n=-\infty}^{\infty} ,\\
    \displaystyle
    \omega \rightarrow \omega_n\equiv \frac{2n}{\beta}\pi .
  \end{array}
  \right.
\label{replace}
\end{eqnarray}
Thus the effective potential at finite temperature
on $R\otimes S^{D-1}$ reads
\begin{eqnarray}
  &&V (\phi) = \frac{\xi_0}{2}R\phi^2 -\frac{\mu_0^2}{2}\phi^2 
    +\frac{\lambda_0}{4!}\phi^4
\nonumber\\
   &&    +\frac{\hbar\lambda_0}{4\beta}\frac{a^{3-D}}{(4\pi)^{(D-1)/2}}
     \Gamma\left(\frac{3-D}{2}\right)
\label{EP:RST} \\
  && \times \int_0^{\phi^2}dm^2\sum_{n=-\infty}^{\infty}
     \frac{\displaystyle \Gamma\left(\frac{D-2}{2}+i\alpha_S\right)
                    \Gamma\left(\frac{D-2}{2}-i\alpha_S\right)}
          {\displaystyle \Gamma\left(\frac{1}{2}+i\alpha_S\right)
                    \Gamma\left(\frac{1}{2}-i\alpha_S\right)},
\nonumber
\end{eqnarray}
with
\begin{equation}
  \alpha_S = \sqrt{f(\omega_n)a^{2}
             +(D-1)(D-2)\xi_0
             -\frac{(D-2)^{2}}{4}}\, ,
\end{equation}
and on $R\otimes H^{D-1}$
\begin{eqnarray}
  V (\phi) & = & \frac{\xi_0}{2}R\phi^2 -\frac{\mu_0^2}{2}\phi^2 
    +\frac{\lambda_0}{4!}\phi^4
 \nonumber\\
   &&   +\frac{\hbar\lambda_0}{4\beta}\frac{a^{3-D}}{(4\pi)^{(D-1)/2}}
     \Gamma\left(\frac{3-D}{2}\right)
\nonumber \\
  && \times \int_0^{\phi^2}dm^2\sum_{n=-\infty}^{\infty}
     \frac{\displaystyle \Gamma\left(\frac{D-2}{2}+\alpha_H\right)}
          {\displaystyle \Gamma\left(\frac{4-D}{2}+\alpha_H\right)},
\label{EP:RHT}
\end{eqnarray}
with
\begin{equation}
  \alpha_H = \sqrt{f(\omega_n)a^{2}
            -(D-1)(D-2)\xi_0
            +\frac{(D-2)^{2}}{4}}\, .
\end{equation}

As is known, the ordinary loop expansion is not always valid for Bose 
fields at finite temperature. Higher order contributions of loop 
expansion contain terms of the order O$(\lambda T^{D-2} /\phi^{2})$. 
At high temperature these terms are not negligible for $D\gtrsim 4$.
The terms proportional to $T^{D-2}$ are described by Ring diagrams.
We resum all the Ring diagrams to improve the loop expansion for 
$D\gtrsim 4$. 
The resummed result is obtained by the following replacement in
the two-point function\cite{FT}
\begin{eqnarray}
  -\mu_0^2 &\rightarrow& \Pi =  -\mu_0^2
\nonumber\\
  && +\pi^{(D-5)/2}
  \Gamma\left(\frac{3-D}{2}\right)
  \zeta(3-D)\frac{\hbar\lambda_0T^{D-2}}{4} .
\label{resum}
\end{eqnarray}
For $D=4$ Eq.(\ref{resum}) reduces to the well-known formula,
\begin{equation}
  -\mu_0^2 \rightarrow \Pi = -\mu_0^2+\frac{\hbar\lambda_0T^{2}}{24}.
\label{resum4}
\end{equation}
We apply the replacement (\ref{resum}) to the $\alpha_S$ and $\alpha_H$ in 
Eqs.(\ref{EP:RST}) and (\ref{EP:RHT}).

\subsection{Renormalization}
The effective potential $V(\phi)$ obtained in the previous section
is divergent in two and four dimensions. To obtain the finite effective 
potential we must renormalize the theory in two and four spacetime 
dimensions. 

First we introduce the renormalization procedure in a flat spacetime. 
The effective potential in D-dimensional flat spacetime is given by
\begin{equation}
  V_0 (\phi) = -\frac{\mu_0^2}{2}\phi^2 
    +\frac{\lambda_0}{4!}\phi^4
    -\frac{\hbar}{2(4\pi)^{D/2}}f_0(\phi),
\label{def:v0}
\end{equation}
with
\begin{equation}
  f_0(\phi) = \lambda_0 \int_0^{\phi^2}dm^2\int \frac{d^D k}{(2\pi)^D} 
  \frac{1}{k^2-\mu_0^2+\frac{\lambda_0}{2}m^2}.
\label{def:f0}
\end{equation}
Integrating over $m^2$ and $k$ in Eq.(\ref{def:f0}), we obtain
\begin{eqnarray}
  f_0(\phi) = \Gamma\left(-\frac{D}{2}\right)
  \left[
    \left(-\mu_0^2+\frac{\lambda_0}{2}\phi^2\right)^{D/2}
    -\left(-\mu_0^2\right)^{D/2}
  \right] .
\end{eqnarray}
It is divergent in two- and four-dimensions.

We introduce the renormalization procedure by imposing the renormalization 
conditions
\begin{equation}
  \left. \frac{\partial^2 V_0}{\partial \phi^2} \right|_{\phi=0}
  \equiv -\mu_r^2 , \hspace{2ex}
  \left. \frac{\partial^4 V_0}{\partial \phi^4} \right|_{\phi=M}
  \equiv \lambda_r ,
\label{def:r}
\end{equation}
where $M$ is the renormalization scale. From these conditions
we obtain the renormalized mass parameter $\mu_r$ and the coupling constant 
$\lambda_r$ ;
\begin{equation}
  \mu_r^2 = \mu_0^2-\frac{\hbar\lambda_0}{2(4\pi)^{D/2}}
            \Gamma\left(1-\frac{D}{2}\right)
            (-\mu_0^2)^{D/2-1} ,
\label{def:mr}
\end{equation}
\begin{eqnarray}
  &&\lambda_r  =  \lambda_0^2-\frac{\hbar\lambda_0^2}{2(4\pi)^{D/2}}
                  \Gamma\left(2-\frac{D}{2}\right)
                  \left(-\mu_0^2+\frac{\lambda_0}{2}M^2\right)^{D/2-2}
\nonumber \\
  && \times \left[
       3+6\lambda_0 M^2 \left(\frac{D}{2}-2\right)
       \left(-\mu_0^2+\frac{\lambda_0}{2}M^2\right)^{-1}
     \right.
\label{def:lr}\\
\nonumber 
  && \left.
       \hspace*{2ex}
       +\lambda_0^2 M^4 \left(\frac{D}{2}-2\right)\left(\frac{D}{2}-3\right)
       \left(-\mu_0^2+\frac{\lambda_0}{2}M^2\right)^{-2}
     \right] .
\end{eqnarray}

Replacing the bare parameters, $\mu_0$ and $\lambda_0$, with the
renormalized ones, $\mu_r$ and $\lambda_r$, we obtain the renormalized 
effective potential 
\begin{eqnarray}
  &&V_0 (\phi)  =  -\frac{\mu_r^2}{2}\phi^2 
    +\frac{\lambda_r}{4!}\phi^4
\nonumber \\
  && -\frac{\hbar}{2(4\pi)^{D/2}}
    \Gamma\left(-\frac{D}{2}\right)
    \left[
      \left(-\mu_r^2+\frac{\lambda_r}{2}\phi^2\right)^{D/2}
      -\left(-\mu_r^2\right)^{D/2}
    \right]
\nonumber \\
  && -\frac{\hbar\lambda_r}{4(4\pi)^{D/2}}\phi^2
    \Gamma\left(1-\frac{D}{2}\right) (-\mu_r)^{D/2-1}
\nonumber \\
  && +\frac{\hbar\lambda_r^2}{2\cdot 4!(4\pi)^{D/2}}\phi^4
                  \Gamma\left(2-\frac{D}{2}\right)
                  \left(-\mu_r^2+\frac{\lambda_r}{2}M^2\right)^{D/2-2}
\nonumber \\
  &&   \hspace*{1ex}\times \left[
       3+6\lambda_r M^2 \left(\frac{D}{2}-2\right)
       \left(-\mu_r^2+\frac{\lambda_r}{2}M^2\right)^{-1}
     \right.
\nonumber \\
  && \left.
       \hspace*{2ex}
       +\lambda_r^2 M^4 \left(\frac{D}{2}-2\right)\left(\frac{D}{2}-3\right)
       \left(-\mu_r^2+\frac{\lambda_r}{2}M^2\right)^{-2}
     \right] .
\label{def:Vr}
\end{eqnarray}
If we take the two-dimensional limit, $D\rightarrow 2$, 
the renormalized potential reduces to
\begin{eqnarray}
  V_0 (\phi)  &=&  -\frac{\mu_r^2}{2}\phi^2 
    +\frac{\lambda_r}{4!}\phi^4
\nonumber \\
 &&   -\frac{\hbar}{8\pi}\left(-\mu_r^2-\frac{\lambda_r}{2}\phi^2\right)
       \ln \left(\frac{-\mu_r^2+\lambda_r\phi^2/2}{-\mu_r^2}\right)
\nonumber \\
  && + \frac{\hbar\lambda_r}{16\pi}\phi^2
       +\frac{\hbar\lambda_r^2\phi^4}{192\pi}\left[
       3\left(-\mu_r^2+\frac{\lambda_r}{2}M^2\right)^{-1}
     \right.
\nonumber\\
  && \hspace*{6ex}
       -6\lambda_r M^2 \left(-\mu_r^2+\frac{\lambda_r}{2}M^2\right)^{-2}
\nonumber \\
  && \left.\hspace*{6ex}
       +2\lambda_r^2 M^4 \left(-\mu_r^2+\frac{\lambda_r}{2}M^2\right)^{-3}
     \right] .
\label{def:Vr2}
\end{eqnarray}
Taking the four-dimensional limit, we obtain the renormalize effective
potential 
\begin{eqnarray}
  V_0 (\phi) & = & -\frac{\mu_r^2}{2}\phi^2 
    +\frac{\lambda_r}{4!}\phi^4
\nonumber \\
  &&  + \frac{\hbar}{64\pi^2}\mu_r^2\left(\mu_r^2-\lambda_r\phi^2\right)
       \ln \left(\frac{-\mu_r^2+\lambda_r\phi^2/2}{-\mu_r^2}\right)
\nonumber \\
  && + \frac{\hbar\lambda_r^2}{256\pi^2}\phi^4
       \ln \left(\frac{-\mu_r^2+\lambda_r\phi^2/2}
       {-\mu_r^2+\lambda_r M^2/2}\right)
\nonumber \\
   &&  + \frac{\hbar\lambda_r}{128\pi^2}\mu_r^2\phi^2
     - \frac{\hbar3\lambda_r^2}{512\pi^2}\phi^4
\nonumber \\
  && - \frac{\hbar\lambda_r^3}{128\pi^2}
       \frac{M^2}{-\mu_r^2+\lambda_r M^2/2}\phi^4
\nonumber \\
  &&   + \frac{\hbar\lambda_r^4}{768\pi^2}
       \frac{M^4}{(-\mu_r^2+\lambda_r M^2/2)^2}\phi^4 .
\label{def:Vr4}
\end{eqnarray}
Thus the finite expression of the effective potential is obtained.

In the two- and the four-dimensional manifolds $R\otimes S^{D}$ 
and $R\otimes H^{D}$ we obtain a finite expression for the effective 
potential by the renormalization conditions (\ref{def:mr}) and 
(\ref{def:lr}).  
To see it we consider the conformal coupled case for simplicity. 
Under the conformal coupling
\begin{equation}
\xi_0 = \frac{D-2}{4(D-1)} ,
\end{equation}
the effective potential (\ref{EP:RS}) on $R\otimes S^{D}$ 
for $D=2$ and $D=4$ reads
\begin{equation}
  V (\phi) = V_0(\phi)+\frac{1}{2a^2}\phi^2 
             +\frac{\hbar}{4} f^D(\phi),
\label{V:rs}
\end{equation}
with
\begin{eqnarray}
  f^{D=2}(\phi)
     &=& -\frac{\lambda_r}{4} \int_0^{\phi^2}dm^2
       \int_{-\infty}^{\infty}\frac{d\omega}{2\pi}
       \frac{1}{\sqrt{\omega^2-\mu_r^2+\lambda_r m^2/2}}
\nonumber \\
     && \hspace*{6ex}\times
        \frac{1}{1-e^{2\pi a \sqrt{\omega^2-\mu_r^2+\lambda_r m^2/2}}} ,
\end{eqnarray}
and
\begin{equation}
  f^{D=4}(\phi)
     = \frac{\lambda_r}{4\pi} \int_0^{\phi^2}dm^2
       \int_{-\infty}^{\infty}\frac{d\omega}{2\pi}
       \frac{\sqrt{\omega^2-\mu_r^2+\lambda_r m^2/2}}
            {1-e^{2\pi a \sqrt{\omega^2-\mu_r^2+\lambda_r m^2/2}}} .
\end{equation}
The additional term $f^D(\phi)$ is obviously convergent.

On $R\otimes H^{D}$ the effective potential (\ref{EP:RH}) for 
for $D=2$ and $D=4$ reduces to
\begin{equation}
  V (\phi) = V_0(\phi)-\frac{1}{2a^2}\phi^2 .
\label{V:rh}
\end{equation}

There is no divergent terms which depend on the scale factor $a$.
The ultra-violet divergence in the effective potential (\ref{V:rs})
and (\ref{V:rh}) is cancelled out by using the same renormalized
parameters (\ref{def:mr}) and (\ref{def:lr}) obtained 
in a flat spacetime. The renormalized effective potential $V(\phi)$
is given by replacing $V_0$ with the one in Eq.(\ref{def:Vr}).

Extension to the finite temperature case is also trivial.
As is well-known, thermal effects do not change the ultra-violet
behavior of the theory. Thus no new divergent terms appear at finite 
temperature (See $V_T(\phi)$ in the next section). All the divergent 
terms are cancelled out by applying the renormalization procedure 
for $T=0$ in a flat spacetime. 

\section{Numerical calculation}
We wish to observe the thermal and curvature effect on the effective 
potential. For this purpose the effective potential is calculated
numerically as a function of the field $\phi$. 

The expressions (\ref{EP:RST}) and (\ref{EP:RHT}) are not useful
for a numerical analysis, since the divergent part is not clearly 
separated. Hence the summation appeared in the expression is not 
convergent. To obtain the finite expression of the effective potential 
we use the following trick \cite{II}.

At finite temperature in a flat spacetime the effective potential
for $\lambda\phi^4$ theory is given by
\begin{eqnarray}
  && V (\phi) = -\frac{\mu_0^2}{2}\phi^2 
    +\frac{\lambda_0}{4!}\phi^4
\nonumber \\
    &&+\frac{\hbar\lambda}{4\beta}
    \int_0^{\phi^2}dm^2\int\frac{d^{D-1}\boldsymbol{k}}{(2\pi)^{D-1}}
    \sum_{n=-\infty}^{\infty}\frac{1}{\omega_n^2+\boldsymbol{k}^2+\Pi
    +\frac{\lambda}{2}m^2},
\nonumber\\
\label{epot:ft}
\end{eqnarray}
where all the ring diagrams is resummed for $D\gtrsim 4$. We replace
$-\mu^2$ to $\Pi$ (\ref{resum}) in the propagator. If the
spacetime dimensions are much smaller than four, for example 
$2 \leq D \leq 3$, we put $\Pi = -\mu^2$. 
We perform the integration over $\boldsymbol{k}$ in Eq.(\ref{epot:ft}) and
find
\begin{eqnarray}
  V (\phi) &=& -\frac{\mu_0^2}{2}\phi^2 
    +\frac{\lambda_0}{4!}\phi^4
\nonumber \\
    &&+\frac{\hbar\lambda}{4\beta}\frac{1}{(4\pi)^{(D-1)/2}}
    \Gamma\left(\frac{3-D}{2}\right)
\label{vt1}
\\
 &&\times\int_0^{\phi^2}dm^2
    \sum_{n=-\infty}^{\infty}
    (\omega_n^2+\Pi+\frac{\lambda}{2}m^2)^{(D-3)/2}.
\nonumber
\end{eqnarray}
The divergent term for $T=0$ is contained in the infinite summation. 

If we perform a summation and integration over $m^2$ and angle
variables first in Eq. (\ref{epot:ft}) and leave the integration
over $k$, we obtain the following expression for the effective potential
\begin{eqnarray}
  V (\phi) &=& -\frac{\mu_0^2}{2}\phi^2 
    +\frac{\lambda_0}{4!}\phi^4
\nonumber \\
    &&-\frac{\hbar}{2}\frac{1}{(4\pi)^{D/2}}
    \Gamma\left(-\frac{D}{2}\right)
    \left[
    (\Pi+\frac{\lambda}{2}\phi^2)^{D/2}
    -\Pi^{D/2}
    \right]
\nonumber\\
    &&+\frac{\hbar}{\beta}\frac{2}{(4\pi)^{(D-1)/2}}
    \frac{1}{\displaystyle \Gamma\left(\frac{D-1}{2}\right)}
\nonumber\\
&& \times
    \int k^{D-2}dk
\ln \frac{1-e^{-\beta\sqrt{k^2+\Pi+\lambda\phi^2/2}}}
                            {1-e^{-\beta\sqrt{k^2+\Pi}}}.
\nonumber \\
&=& V_0(\phi)+V_T(\phi) ,
\label{vt2}
\end{eqnarray}
where $V_0(\phi)$ is the effective potential (\ref{def:v0}) for $T=0$ in a 
flat spacetime. The divergent term at $T=0$ is included in $V(0)$. The 
divergence is cancelled out after the renormalization discussed in the 
previous section. $V_T(\phi)$ is finite in the spacetime dimensions 
$2\leq D < 5$.

Comparing Eq.(\ref{vt1}) with Eq.(\ref{vt2}), we find the relationship
\begin{eqnarray}
    &&-\frac{\hbar}{2}\frac{1}{(4\pi)^{D/2}}
    \Gamma\left(-\frac{D}{2}\right)
    \left[
    (\Pi+\frac{\lambda}{2}\phi^2)^{D/2}
    -\Pi^{D/2}
    \right]
\nonumber \\
    &&+\frac{\hbar}{\beta}\frac{2}{(4\pi)^{(D-1)/2}}
    \frac{1}{\displaystyle \Gamma\left(\frac{D-1}{2}\right)}
\nonumber \\
   &&\hspace*{2ex}\times
 \int k^{D-2}dk
 \ln \frac{1-e^{-\beta\sqrt{k^2+\Pi+\lambda\phi^2/2}}}
                            {1-e^{-\beta\sqrt{k^2+\Pi}}}
\\
\nonumber
    &&=\frac{\hbar\lambda}{4\beta}\frac{1}{(4\pi)^{(D-1)/2}}
    \Gamma\left(\frac{3-D}{2}\right)
\nonumber \\
   &&\hspace*{2ex}\times
    \int_0^{\phi^2}dm^2
    \sum_{n=-\infty}^{\infty}
    (\omega_n^2+\Pi+\frac{\lambda}{2}m^2)^{(D-3)/2}.
\label{trick}
\end{eqnarray}

Inserting Eq.(\ref{trick}) into Eqs.(\ref{EP:RST}) and (\ref{EP:RHT}), 
we obtain the finite expression for the effective potential on 
$R\otimes S^{D}$
\begin{eqnarray}
  &&V(\phi) = V_0(\phi)+\frac{\xi}{2}R\phi^2
\nonumber \\
    &&+\frac{\hbar}{\beta}\frac{2}{(4\pi)^{(D-1)/2}}
    \frac{1}{\displaystyle \Gamma\left(\frac{D-1}{2}\right)}
\nonumber \\
   &&\hspace*{2ex}\times
    \int k^{D-2}dk
    \ln \frac{1-e^{-\beta\sqrt{k^2+\Pi+\lambda\phi^2/2}}}
                            {1-e^{-\beta\sqrt{k^2+\Pi}}}
\nonumber \\
    &&-\frac{\hbar\lambda}{4\beta}\frac{1}{(4\pi)^{(D-1)/2}}
    \Gamma\left(\frac{3-D}{2}\right)
\nonumber \\
   &&\hspace*{2ex}\times
    \int_0^{\phi^2}dm^2
    \sum_{n=-\infty}^{\infty}
    \left[\left(\omega_n^2+\Pi+\frac{\lambda}{2}m^2\right)^{(D-3)/2}
    \vphantom{\frac{\displaystyle \Gamma\left(\frac{D-2}{2}+i\alpha_S\right)
                    \Gamma\left(\frac{D-2}{2}-i\alpha_S\right)}
         {\displaystyle \Gamma\left(\frac{1}{2}+i\alpha_S\right)
                    \Gamma\left(\frac{1}{2}-i\alpha_S\right)}
    } \right.
\nonumber \\
    && \left.\hspace*{2ex}
      -a^{3-D}\frac{\displaystyle \Gamma\left(\frac{D-2}{2}+i\alpha_S\right)
                    \Gamma\left(\frac{D-2}{2}-i\alpha_S\right)}
          {\displaystyle \Gamma\left(\frac{1}{2}+i\alpha_S\right)
                    \Gamma\left(\frac{1}{2}-i\alpha_S\right)}\right],
\label{nc:vs}
\end{eqnarray}
and on $R\otimes H^{D}$
\begin{eqnarray}
  &&V(\phi) = V_0(\phi)+\frac{\xi}{2}R\phi^2
\nonumber \\
    &&+\frac{\hbar}{\beta}\frac{2}{(4\pi)^{(D-1)/2}}
    \frac{1}{\displaystyle \Gamma\left(\frac{D-1}{2}\right)}
\nonumber \\
   &&\hspace*{2ex}\times
    \int k^{D-2}dk
 \ln \frac{1-e^{-\beta\sqrt{k^2+\Pi+\lambda\phi^2/2}}}
                            {1-e^{-\beta\sqrt{k^2+\Pi}}}
\nonumber \\
    &&-\frac{\hbar\lambda}{4\beta}\frac{1}{(4\pi)^{(D-1)/2}}
    \Gamma\left(\frac{3-D}{2}\right)
\nonumber \\
   &&\hspace*{2ex}\times
    \int_0^{\phi^2}dm^2
    \sum_{n=-\infty}^{\infty}
    \left[\left(\omega_n^2+\Pi+\frac{\lambda}{2}m^2\right)^{(D-3)/2}
    \vphantom{\frac{\displaystyle \Gamma\left(\frac{D-2}{2}+\alpha_H\right)}
          {\displaystyle \Gamma\left(\frac{4-D}{2}+\alpha_H\right)}
    } \right.
\nonumber \\
    && \left.\hspace*{2ex}
      -a^{3-D}\frac{\displaystyle \Gamma\left(\frac{D-2}{2}+\alpha_H\right)}
          {\displaystyle \Gamma\left(\frac{4-D}{2}+\alpha_H\right)}\right] .
\label{nc:vh}
\end{eqnarray}

The renormalized effective potential $V(\phi)$ is given by replacing 
$V_0$ with the one in Eq.(\ref{def:Vr}). It should be noted that we 
replace $-\mu^2$ with $\Pi$ in the second line of Eq. (\ref{def:Vr})
for $D\gtrsim 4$ to improve the loop expansion at high temperature. 
By using the expressions (\ref{nc:vs}) and (\ref{nc:vh}) we show some 
characteristic behaviors of the effective potential near the critical 
point where the phase transition takes place.
The effective potential develops a non-vanishing imaginary part in
the present renormalization conditions.
Below we draw the real part of the effective potential.
We take the natural unit and put $\hbar=1$. All the mass scale is 
normalized by the renormalization scale $M$.

\begin{figure}
\includegraphics{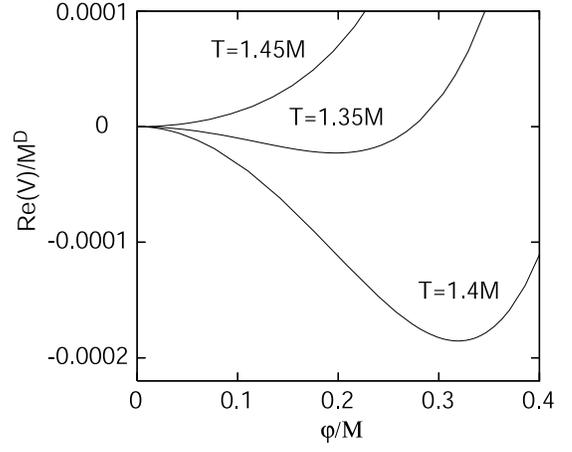}
\caption{\label{vTD35mu01}Behavior of the effective potential for $\mu^2=0.1M^2$, $\lambda=M^{4-D}$ and $D=3.5$ in a flat spacetime as $T$ varies.}
\end{figure}
It is well-known that the $Z_2$ symmetry under the transformation 
(\ref{def:Vr}) is broken spontaneously for $\mu \geq 0$ in an infinite 
volume flat spacetime (i.e., Minkowski spacetime) at $T=0$.
As is seen in Fig. \ref{vTD35mu01}, the symmetry broken spontaneously 
is restored at high temperature. If the temperature is no less then a 
critical value $T_c$, the minimum of the effective potential locates 
at $\phi=0$. The ground state keeps the $Z_2$ symmetry under the 
transformation (\ref{def:disc}). There is a second order phase transition 
as $T$ is increased for $D=3.5$.

\begin{figure}
\includegraphics{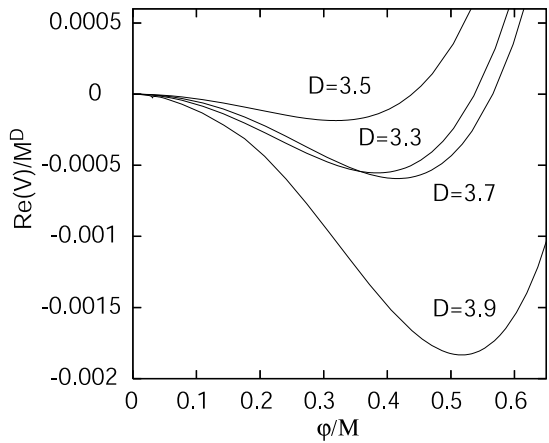}
\caption{\label{vTmu01T235a}Behavior of the effective potential for $T=1.35M$ in a flat spacetime at $D=3.3, 3.5, 3.7, 3.9$ for $\mu^2=0.1M^2$ and $\lambda=M^{4-D}$.}
\end{figure}

\begin{figure}
\includegraphics{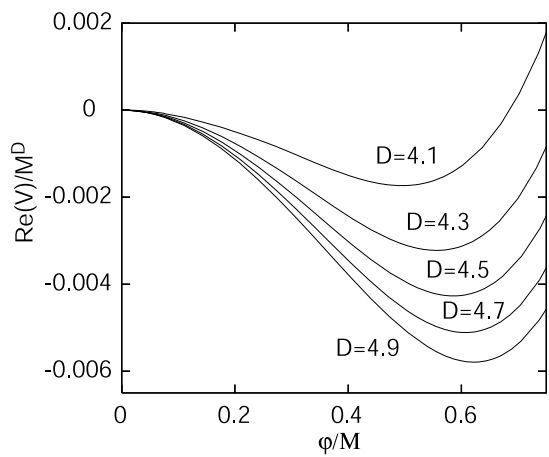}
\caption{\label{vTmu01T235b}Behavior of the effective potential for $T=1.35M$ in a flat spacetime at $D=4.1, 4.3, 4.5, 4.7, 4.9$ for $\mu^2=0.1M^2$ and $\lambda=M^{4-D}$.}
\end{figure}

In Fig. \ref{vTmu01T235a} and \ref{vTmu01T235b} a $D$ dependence of 
the effective potential is presented with the temperature fixed at
$T=2.35M$. To draw the figures we use the resumed propagator. The dependence 
becomes larger near the four spacetime dimensions. It seems that the 
critical temperature has a gap at an integer dimensions.

\begin{figure*} 
\begin{minipage}{8cm}
\includegraphics{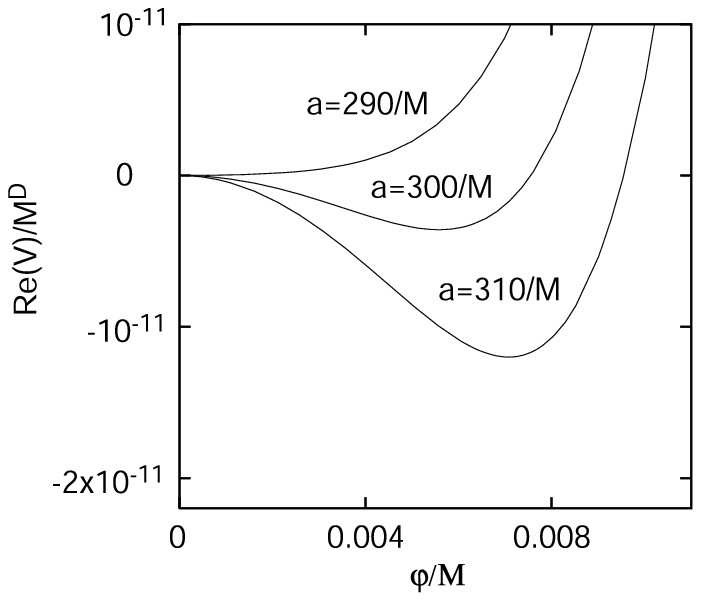}

(a) $\mu^2=0$, $\lambda=M^{4-D}$, $T=0.01M < T_c$
\end{minipage}
\begin{minipage}{8cm}
\includegraphics{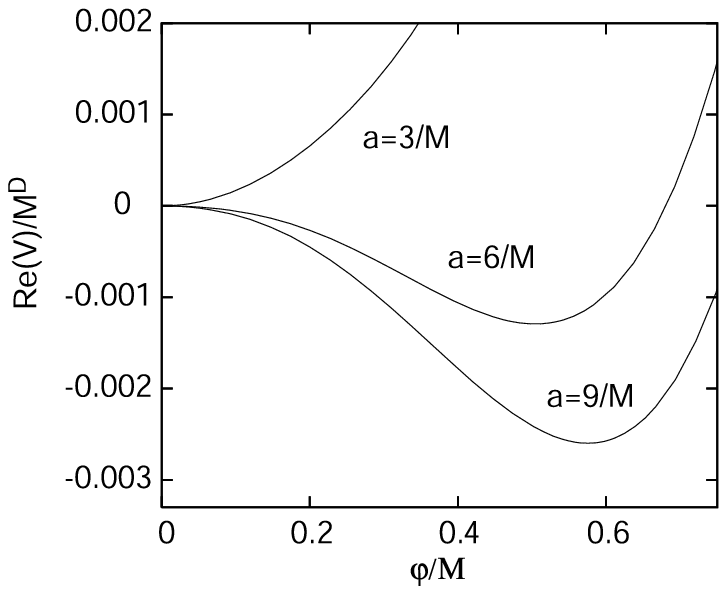}

(b) $\mu^2=0.1M^2$, $\lambda=M^{4-D}$, $T=1.0M < T_c$
\end{minipage}
\caption{\label{fig:vSTconfmu01} Behavior of the effective potential for a conformally coupled scalar field $\xi=(D-2)/(4D-4)$ in $R\otimes S^{2.5}$ as $a$ varies.}
\end{figure*}

\begin{figure*}
\begin{minipage}{8cm}
\includegraphics{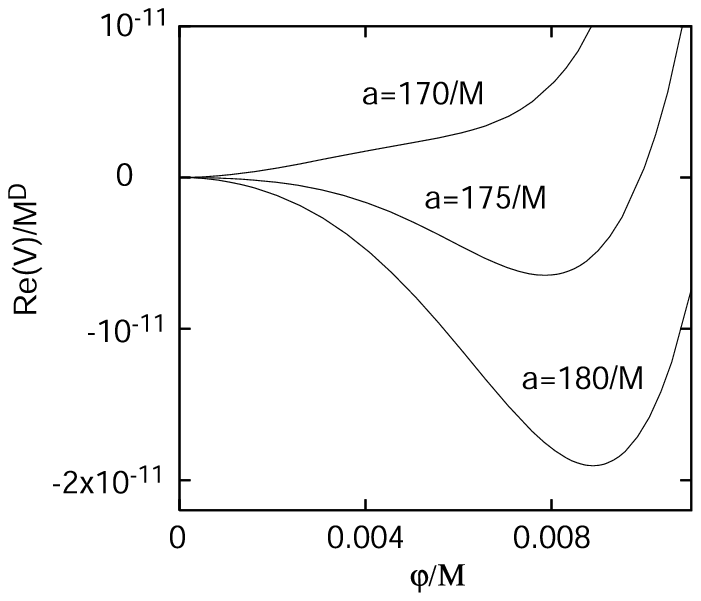}

(a) $\mu^2=0$, $\lambda=M^{4-D}$, $T=0.01M < T_c$
\end{minipage}
\begin{minipage}{8cm}
\includegraphics{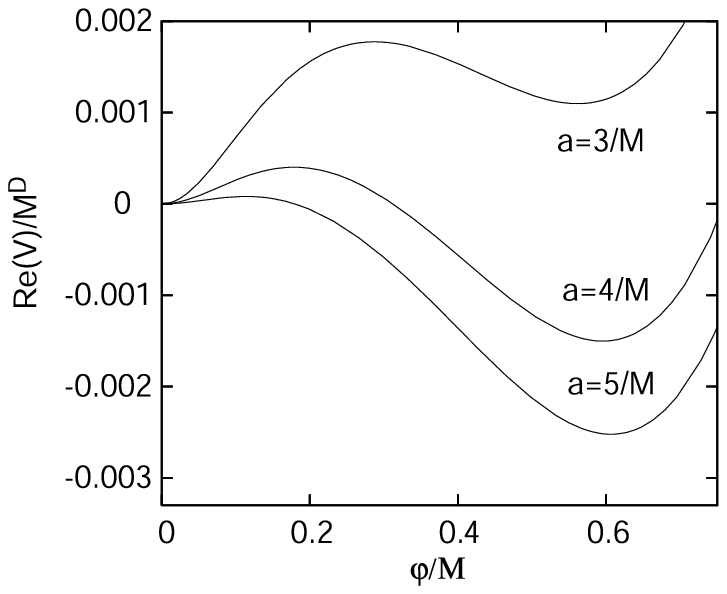}

(b) $\mu^2=0.1M^2$, $\lambda=M^{4-D}$, $T=1.0M < T_c$
\end{minipage}
\caption{\label{fig:vSTminimu01} Behavior of the effective potential for a minimally coupled scalar field $\xi=0$ in $R\otimes S^{2.5}$ as $a$ varies.}
\end{figure*}

We observe the behavior of the effective potential in $R\otimes S^{D-1}$ 
at finite temperature. 
It is found in Ref. \cite{O'Connor:1983fq} that a positive curvature
suppress the symmetry breaking for $T=0$ in the Einstein universe. 
To see the curvature effect at finite temperature, we take a
temperature lower than the critical one and calculate the effective 
potential in $R\otimes S^{D-1}$. The behaviors of the effective is 
illustrated for a conformally coupled
scalar field, $\xi=(D-2)/(4D-4)$, and for a minimally coupled one, 
$\xi=0$ in Figs. \ref{fig:vSTconfmu01} and \ref{fig:vSTminimu01}, 
respectively.
As is seen in the figures, a positive curvature restores the symmetry. 
The ground state is always symmetric under the transformation 
(\ref{def:disc}) for $T\geq T_c$ in $R\otimes S^{D-1}$. For $\mu=0$ the 
spontaneous symmetry breaking is induced by only a radiative correction
\cite{Col}. 
The broken symmetry is restored by a larger scale factor, in other words, 
a smaller curvature. 
There is the first order phase transition as $a$ is decreased with $T$ 
kept small and fixed.

The curvature effect to the symmetry breaking comes from a curvature 
dependence of the covariant derivative and a coupling $\xi$ between the 
scalar field and the gravitational field. 
We clearly observe in Fig. \ref{fig:vSTminimu01}, that the curvature 
effect can restore the broken symmetry without a coupling $\xi$.
The curvature effect for $\xi=0$ is smaller than the one for a conformally 
coupled case.

\begin{figure*}
\begin{minipage}{8cm}
\includegraphics{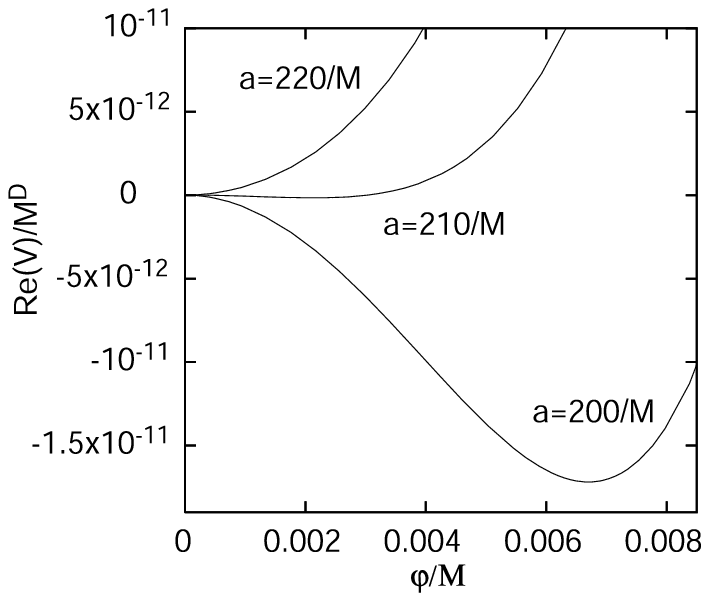}

(a) $\mu^2=0$, $\lambda=M^{4-D}$, $T=0.02M > T_c$
\end{minipage}
\begin{minipage}{8cm}
\includegraphics{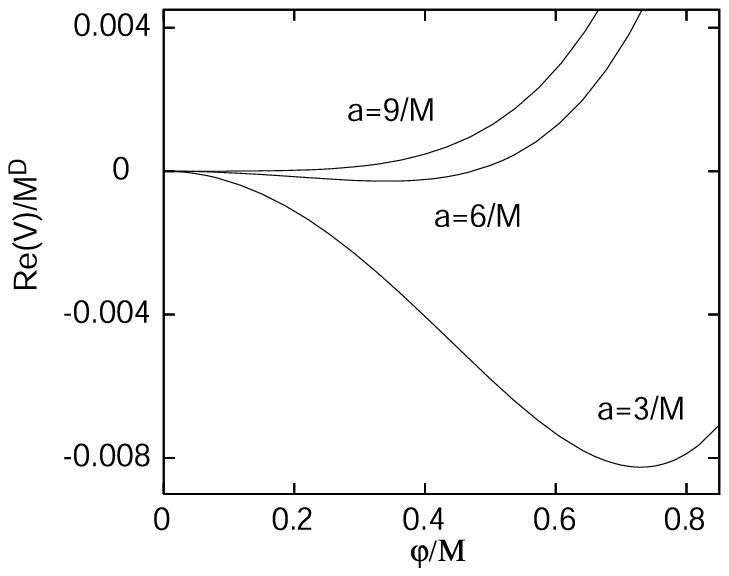}

(b) $\mu^2=0.1M^2$, $\lambda=M^{4-D}$, $T=1.5M > T_c$
\end{minipage}
\caption{\label{fig:vHTconfmu01} Behavior of the effective potential for a conformally coupled scalar field $\xi=(D-2)/(4D-4)$ in $R\otimes H^{2.5}$ as $a$ varies.}
\end{figure*}

\begin{figure*}
\begin{minipage}{8cm}
\includegraphics{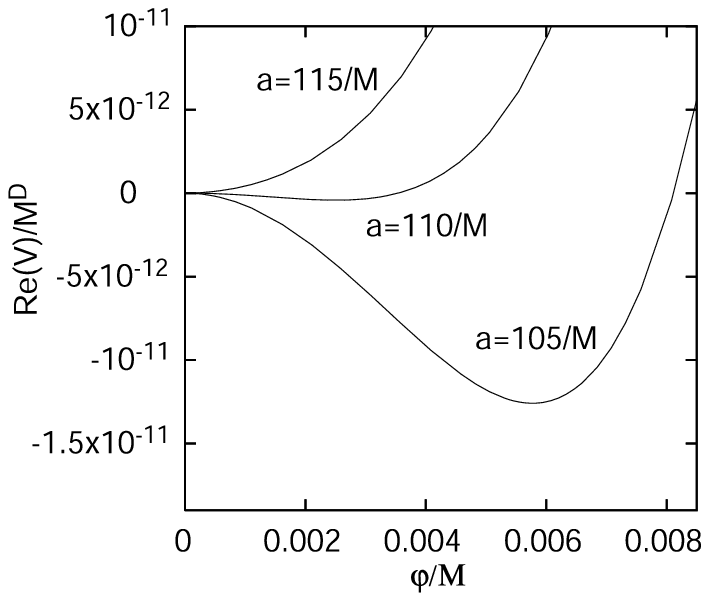}

(a) $\mu^2=0$, $\lambda=M^{4-D}$, $T=0.02M > T_c$
\end{minipage}
\begin{minipage}{8cm}
\includegraphics{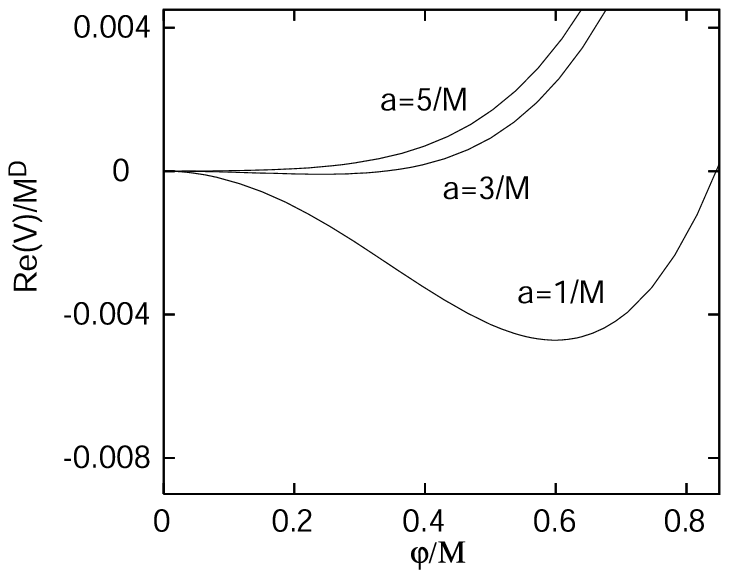}

(b) $\mu^2=0.1M^2$, $\lambda=M^{4-D}$, $T=1.5M > T_c$
\end{minipage}
\caption{\label{fig:vHTminimu01} Behavior of the effective potential for a minimally coupled scalar field $\xi=0$ in $R\otimes H^{2.5}$ as $a$ varies.}
\end{figure*}

Next we study the curvature effect in the negative curvature space 
$R\otimes H^{D-1}$ at finite temperature. 
We fix the temperature above the critical one and see whether the
$Z_2$ symmetry is broken down in an environment of the small
scale factor $a$. In Figs. \ref{fig:vHTconfmu01} and \ref{fig:vHTminimu01}
we plot the typical behavior of the effective potential at $D=3.5$ as
the scale factor $a$ varies. As is seen in the figures, we observe that 
there is a second order phase transition and the $Z_2$ symmetry
is broken down as $a$ is decreased.
In the negative curvature case a larger scale factor is enough to break 
the symmetry for $\mu=0$. We see in Fig. \ref{fig:vHTminimu01}, a negative 
curvature breaks the symmetry without a coupling $\xi$.

\section{Concluding Remarks}
We have investigated the behavior of the effective potential at finite
temperature and curvature in arbitrary dimensions $(2\leq D < 5)$.
We discuss the curvature and thermal effects to the effective potential
for both a minimally coupled and conformally coupled scalar fields. 

Starting from the theory with broken $Z_2$ symmetry at $T=0$ and
$a \rightarrow \infty$, we calculate the renormalized effective 
potential for finite $T$ and $a$. The broken $Z_2$ symmetry is restored
at a certain critical temperature and scale factor in $R\otimes S^{D-1}$. 
Above the critical temperature the restored symmetry is broken down again
at a certain critical scale factor in $R\otimes H^{D-1}$.
The phase transition from the broken phase to the symmetric phase
is either of the first order or the second order for $D=3.5$. 
The critical temperature and scale factor depend on the parameters of 
the theory and $D$.

The effective potential divelops a non-vanishing imaginary part.
If we take the other renormaliztion conditions,
\begin{equation}
  \left. \frac{\partial^2 V_0}{\partial \phi^2} \right|_{\phi=\sqrt{2/\lambda}\ \mu}
  \equiv -\mu_r^2 , \hspace{2ex}
  \left. \frac{\partial^4 V_0}{\partial \phi^4} \right|_{\phi=M}
  \equiv \lambda_r ,
\label{def:rn}
\end{equation}
we can define a real effective potential for $D\geq 4$. 

Although the present work is restricted to the calculation of the
effective potential, we are interested in applying our result to the 
full analysis of the phase structure and physical problems.
A consequence of symmetry breaking may be found to study critical 
phenomena in the early stage of universe. It gives rise to a possibility 
that some cosmological observable show sings of symmetry breaking.
The phase transition may affect the evolution of the spacetime structure. 
The spacetime structure depends on the ground state of the system trough
the expectation value of the stress tensor. This dependence may change 
the minimum of the effective potential. Thus the spacetime evolution may 
affect the symmetry breaking. To obtain the contribution of this back 
reaction we must solve the Einstein equation and the gap equation 
simultaneously \cite{BA, AS}. 

The thermal effect may be significantly stronger in most of realistic 
situations at the early stage of the universe. The curvature effect
may cause a nonstatic field configurations. Decreasing the temperature, 
spontaneous symmetry breaking may occur from the negative curvature place. 
There is a possibility to observe the combined effect of the temperature 
and curvature is a fluctuation of some fields. However, we cannot deal with 
the nonstatic configurations in the effective potential approach. 
For summing up contributions from different region we will need a new idea. 

\begin{acknowledgments}
The authors would like to thank T.~Fujihara and D.~Kimura for
useful discussions. We also thank the Yukawa Institute for Theoretical 
Physics at Kyoto University. Discussions during the YITP workshop 
YITP-W-04-07 on "Thermal Quantum Field Theories and Their Applications" 
were useful to complete this work. 
\end{acknowledgments}

\end{document}